# Online Evaluations for Everyone: Mr. DLib's Living Lab for Scholarly Recommendations

Joeran Beel[1,2], Andrew Collins[1], Oliver Kopp[3], Linus W. Dietz[4], and Petr Knoth[5]

[1] Trinity College Dublin, School of Computer Science & Statistics, ADAPT Centre, Ireland
[2] National Institute of Informatics Tokyo, Digital Content and Media Sciences Division, Japan
[3] IPVS, University of Stuttgart, Germany
[4] Technical University of Munich, Department of Informatics, Garching, Germany
[5] The Open University, Knowledge Media institute, United Kingdom

`beelj@tcd.ie, ancollin@tcd.ie, kopp@informatik.uni-stuttgart.de,`
`linus.dietz@tum.de, petr.knoth@open.ac.uk`

**Abstract.** We introduce the first 'living lab' for scholarly recommender systems. This lab allows recommender-system researchers to conduct online evaluations of their novel algorithms for scholarly recommendations, i.e., recommendations for research papers, citations, conferences, research grants, etc. Recommendations are delivered through the living lab's API to platforms such as reference management software and digital libraries. The living lab is built on top of the recommender-system as-a-service Mr. DLib. Current partners are the reference management software JabRef and the CORE research team. We present the architecture of Mr. DLib's living lab as well as usage statistics on the first sixteen months of operating it. During this time, 1,826,643 recommendations were delivered with an average click-through rate of 0.21%.

## 1 Introduction

'Living labs' for recommender systems enable researchers to evaluate their recommendation algorithms with real users in realistic scenarios. Such living labs – sometimes also called 'Evaluations-as-a-Service' [1–3] – are usually built on top of production recommender systems in real-world platforms such as news websites [4]. Via an API, external researchers can 'plug-in' their experimental recommender systems to the living lab. When recommendations for users of the platform are needed, the living lab sends a request to the researcher's experimental recommender system. This system then returns a list of recommendations that are displayed to the user. The user's actions (clicks, downloads, purchases, etc.) are logged and can be used to evaluate the recommendation algorithms' effectiveness.

Living labs are available in information retrieval and for many recommender-system domains, particularly news [4–6], and they attracted dedicated workshops [7]. There is also work on living labs in the context of search and browsing behavior in digital

This publication has emanated from research conducted with the financial support of Science Foundation Ireland (SFI) under Grant Number 13/RC/2106. We are further grateful for the support received by Samuel Pearce and Siddharth Dinesh.

libraries [8]. However, to the best of our knowledge, there are no living labs for scholarly recommendations, i.e., recommendations for research articles [9,10], citations [11,12], conferences [13,14], reviewers [15,16], quotes [17], research grants, or collaborators [18]. Consequently, researchers in the field of scholarly recommender systems predominately rely on offline evaluations, which tend to be poor predictors of how algorithms will perform in a production recommender system [19,20].

In this paper, we present the first living lab for scholarly recommendations, built on top of *Mr. DLib*, a scholarly recommendations-as-a-service provider [21,22]. Mr. DLib's main feature is to provide third parties such as digital libraries with recommendations for their users. This way, digital libraries do not need to maintain their own recommender system, which would usually be costly and require advanced skills in machine learning and recommender systems. So far, Mr. DLib relied only on its own recommender system to generate recommendations [21,22]. The system was not open to external researchers. The newly added living lab opens Mr. Lib and provides an environment for any researcher in the field of scholarly recommendations to evaluate novel recommendation algorithms with real users in addition to, or instead of, conducting offline evaluations.

## 2 Mr. DLib's Scholarly Living Lab

Mr. DLib's living lab is open for two types of partners. First, platform operators, who want to provide their users with scholarly recommendations. Second, research partners, who want to evaluate their novel scholarly recommendation algorithms with real users. The current platform partner of Mr. DLib is the reference-management software JabRef [23,24]. The current research partner of Mr. DLib is CORE [25–27]. Mr. DLib acts as an intermediate between these partners. Mr. DLib also operates its own internal recommendation engine, which applies content-based filtering with terms, key-phrases, and word embeddings as well as stereotype and most-popular recommendations [22,28]. Thus, Mr. DLib's internal recommendation engine establishes a baseline for research partners to compare their novel algorithms against.

The workflow of Mr. DLib's living lab is illustrated in **Fig. 1**: (1) A JabRef user selects a source article in the list, and then selects the "Related Articles" tab; JabRef sends a request to Mr. DLib's API. The request comprises of the selected article's title. Mr. DLib's API accepts the request, and its A/B engine randomly forwards the request either to (2a) Mr. DLib's internal recommender system or (2b) to CORE's recommender system. CORE or Mr. DLib's internal recommender system creates a list of recommendations and (3) returns them to JabRef, which displays them to the user. (4) When a user clicks a recommendation, a notification it sent to Mr. DLib for evaluation purposes.

While, currently, Mr. DLib only has one research and one platform partner, there will potentially be numerous such partners in the future. Mr. DLib's living lab is open to any research partner whose experimental recommender system recommends scholarly items; is available through a REST API; accepts a string as input (typically a source article's title); and returns a list of related-articles including URLs to web pages on which the recommended articles can be downloaded, preferably open access. Also, recommendations must be returned within less than 2 seconds.

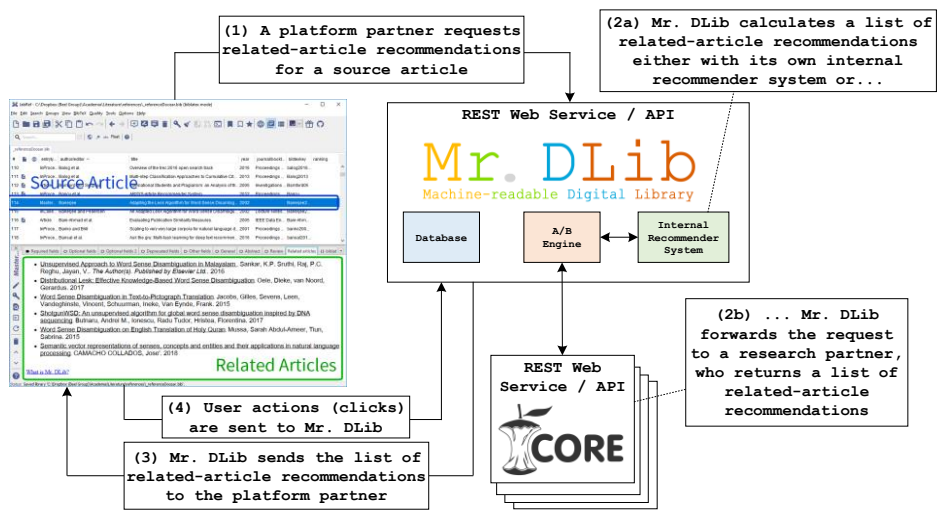

**Fig. 1.** Illustration of the recommendation process

All data on Mr. DLib's recommendations is available publicly [29]. This data can be used to replicate our calculations and perform additional analyses. JabRef's client software, including the recommender system, can be downloaded at http://jabref.org. Source code of the API is available on http://mr-dlib.org.

## 3 Usage Statistics

Mr. DLib started its general recommendation service in 2016 [21] and its living lab in June 2017. The living lab was integrated first in a beta version of JabRef. During the beta phase (until September 2017), JabRef sent around 4,200 requests per month to Mr. DLib (**Fig. 2**). For each request, Mr. DLib returned typically 6 recommendations (25k recommendations in total), whereas between 20% to 30% of the recommendations were generated by CORE, and the remaining by Mr. DLib's internal recommendation engine. Click-through rate (**CTR**) on the recommendations decreased from 0.76% in June to 0.34% in September (**Fig. 2**). After the beta phase, i.e., from October 2017 on, the number of delivered recommendations increased to around 150k per month, again with 20% to 30% of the recommendations generated by CORE. The overall click-through rate decreased to around 0.18% but remained stable until today.

We can only speculate why click-through rate decreased during the beta phase and decreased again in the stable version. Possibly, beta users are more curious than regular users. Maybe users generally are more curious in the beginning when a new feature is released. Maybe, recommendations worsen over time, or were simply not as good as users expected and hence users lose interest. However, we made the observation that

CTR decreases over time also on Mr. DLib's other partner platforms that do not participate in the living lab [22,28], as well as in other recommender systems [30].

Interestingly, click-through rates for both CORE and Mr. DLib's internal recommendation engine are almost identical over the entire data collection period. Both systems mostly use Apache Lucene for their recommendation engine, yet there are notable differences in the algorithms and document corpora. We will not elaborate further on the implementations but refer the interested reader to [22,27,28]. The interesting point here is that two separately implemented recommender systems perform almost identically. It is also interesting that the click-through rate in the reference management software JabRef (0.18%) is quite similar to the click-through rate in the social-science repository Sowiport [28,31–33], although the two platforms differ notably.

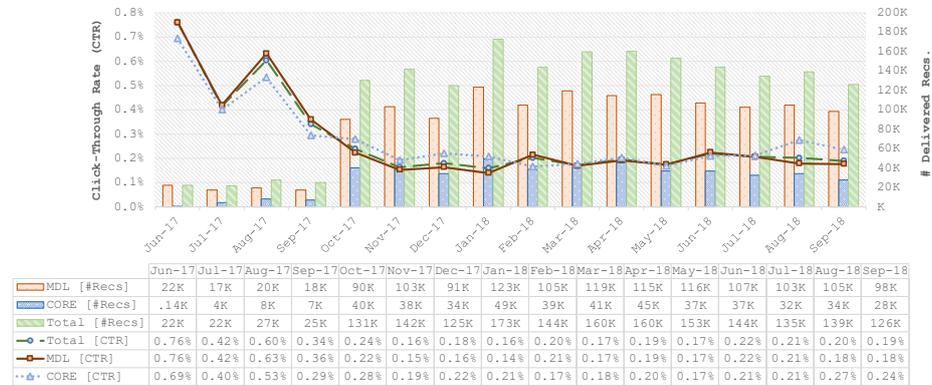

|  | Jun-17 | Jul-17 | Aug-17 | Sep-17 | Oct-17 | Nov-17 | Dec-17 | Jan-18 | Feb-18 | Mar-18 | Apr-18 | May-18 | Jun-18 | Jul-18 | Aug-18 | Sep-18 |
|---|---|---|---|---|---|---|---|---|---|---|---|---|---|---|---|---|
| MDL [#Recs] | 22K | 17K | 20K | 18K | 90K | 103K | 91K | 123K | 105K | 119K | 115K | 116K | 107K | 103K | 105K | 98K |
| CORE [#Recs] | .14K | 4K | 8K | 7K | 40K | 38K | 34K | 49K | 39K | 41K | 45K | 37K | 37K | 32K | 34K | 28K |
| Total [#Recs] | 22K | 22K | 27K | 25K | 131K | 142K | 125K | 173K | 144K | 160K | 153K | 144K | 135K | 139K | 126K |
| Total [CTR] | 0.76% | 0.42% | 0.60% | 0.34% | 0.24% | 0.16% | 0.18% | 0.16% | 0.20% | 0.17% | 0.19% | 0.17% | 0.22% | 0.21% | 0.20% | 0.19% |
| MDL [CTR] | 0.76% | 0.42% | 0.63% | 0.36% | 0.22% | 0.15% | 0.16% | 0.14% | 0.21% | 0.17% | 0.19% | 0.17% | 0.22% | 0.21% | 0.18% | 0.18% |
| CORE [CTR] | 0.69% | 0.40% | 0.53% | 0.29% | 0.28% | 0.19% | 0.22% | 0.21% | 0.17% | 0.18% | 0.20% | 0.17% | 0.21% | 0.21% | 0.27% | 0.24% |

**Fig. 2.** Click-through rate (CTR) and # of delivered recommendation in JabRef for Mr. DLib's (MDL) and CORE's recommendation engine and in total.

## 4 Future Work

In the long-run, we hope to provide a platform to the information retrieval, digital library, and recommender systems community that helps conducting more reproducible and robust research in real-world scenarios [34,35]. To achieve this, we plan to add more partners on both sides – platform partners who provide access to real users, and research partners who evaluate their novel algorithms via the living lab. We also aim for personalized recommendations in addition to the current focus on related-article recommendations. We will also enable the recommendation of other scholarly items such as research grants, or research collaborators. We also plan to develop a more automatic process for the integration of partners, with standard protocols and data formats, and pre-implemented clients, to ease the process. Another major challenge in the future will be to select the best algorithms for each platform partner [36].